\documentclass[aps,twocolumn]{revtex4}

\usepackage{graphicx,epsfig}

\def\duzomniejsze{<\kern-.7mm<}
\def\duzowieksze{>\kern-.7mm>}

\def\textbf#1{{\bf #1}}
\def\be{\begin{equation}}
\def\ee{\end{equation}}
\def\ben{\begin{eqnarray}}
\def\een{\end{eqnarray}}
\def\eea{\end{array}}
\def\bea{\begin{array}}
\newcommand{\bei}{\begin{itemize}}
\newcommand{\eei}{\end{itemize}}
\newcommand{\bee}{\begin{enumerate}}
\newcommand{\eee}{\end{enumerate}}

\newtheorem{theorem}{Theorem}
\newtheorem{definition}{Definition}

\def\blacksquare{\vrule height 4pt width 3pt depth2pt}

\def\dt#1{{{\kern -.0mm\rm d}}#1\,}

\def\tr{{\rm Tr}}

\def\>{\rangle}
\def\<{\langle}
\def\blacksquare{\vrule height 4pt width 3pt depth2pt}

\def\ot{\otimes}

\begin{document}

\title{Simplifying monotonicity conditions for entanglement measures}

\author{Micha\l{} Horodecki}

\affiliation{Institute of Theoretical Physics and Astrophysics,
University of Gda\'nsk, 80--952 Gda\'nsk, Poland
}

\begin{abstract}
We show that for a convex function the following, rather modest conditions, 
are equivalent to monotonicity under local operations and 
classical communication. The conditions are: 
1)invariance under local unitaries, 2) invariance 
under adding local ancilla in arbitrary state 3) on mixtures 
of states possessing local orthogonal flags the function is equal to its average.
The result holds for multipartite systems. It is intriguing 
that the obtained conditions are equalities. The only inequality is 
hidden in the condition of convexity. 
\end{abstract}

\pacs{Pacs Numbers: 03.65.-w}
\maketitle

A basic condition for a function to quantify entanglement is that of nonincreasing 
under local operations and classical communication (LOCC) \cite{BDSW1996,PlenioVedral1998,
Vidal-mon2000} (see \cite{Michal2001} for review). The condition called LOCC monotonicity is 
usually not simple to prove for candidates for entanglement measures. 
The purpose of the paper is to derive conditions equivalent to LOCC monotonicity 
for convex functions. In other words, we consider a convex function $f$
and ask what conditions it should satisfy, to 
be monotone under LOCC. Surprisingly, we obtain 
that the conditions are rather modest. They are  the following: 
1)invariance under local unitaries, 2) invariance 
under adding local ancilla in standard pure state 3) on mixtures 
of states possessing local orthogonal flags the function is equal to its average:
\be
f(\sum_ip_i \rho_i\ot |i\>\<i|) = \sum_ip_i f( \rho_i\ot |i\>\<i|)
\ee
The last condition can be called affinity on locally orthogonal states.
For convex functions, one way inequality follows from convexity. 
It is rather intuitive that the condition 3) is necessary 
for LOCC monotonicity. However it is rather surprising that it is also 
sufficient together with rather modest  conditions 1) and 2). 
Our proofs will be carried out for bipartite states,
however they immediately generalize to multipartite case.

To begin with, let us state more precisely what we mean by monotonicity under 
LOCC. A possible formulation is that for any quantum operation 
$\Lambda$ that can be carried out by means of 
local operations and classical communication we have 
\be
f(\rho_{AB})\geq f(\Lambda(\rho_{AB}))
\label{eq:monweak}
\ee
If we treat LOCC operation as measurement with outcomes $i$,
we can rewrite the condition
\be
f(\rho_{AB})\geq  f(\sum_ip_i\sigma_{AB}^i)
\ee
where $p_i$ are probabilities of outcomes, and $\sigma_{AB}^i$ 
is state  given outcome $i$ was obtained. 

One also considers stronger monotonicity condition, which we will adopt in this paper:
\begin{definition}
\label{def:mon}
A function $f$ is LOCC monotone iff it satisfies the following condition
\be
f(\rho_{AB})\geq \sum_ip_i f(\sigma_{AB}^i)
\ee
where $i$ are outcomes of the LOCC operation, $p_i$ are probabilities of 
outcomes, and $\sigma_{AB}^i$  is state  given outcome $i$ was obtained. 
\end{definition}

We will need the following result of Vidal \cite{Vidal-mon2000}:
\begin{theorem} 
\label{thm:vidal}
A convex function $f$ is LOCC monotone in the sense of Def. \ref{def:mon} 
if and only if it does not increase under 
\bei
\item[a)] adding local ancilla 
\be
f(\rho_{AB}\ot \sigma_X)\leq f(\rho_{AB}),\quad X=A',B' 
\ee
\item[b)] local partial trace 
\be
f(\rho_{AB})\leq f(\rho_{ABX})
\ee
\item[c)] local unitaries 
\item[d)] local von Neumann measurements (not necessarily complete), 
\be
f(\rho_{AB})\geq \sum_i p_i f(\sigma_{AB}^i)
\ee
where $\sigma_{AB}^i$ is state after obtaining outcome $i$,
and $p_i$ is probability of such outcome 
\eei
\end{theorem}

{\bf Remark.} From the proof in \cite{Vidal-mon2000} it is easy to see that 
the above theorem works also for multipartite systems. 

We are now in position to state and prove our new conditions 
equivalent to LOCC monotonicity for convex functions. 

\begin{theorem}
\label{thm:1}
For a convex function $f$ does not increase under  LOCC if and only if 
\bee
\item $f$ is invariant under local unitary operations
\be
f(U_A\ot U_B \rho_{AB} U_A^\dagger\ot U_B^\dagger)= f(\rho_{AB})
\ee
\item $f$ is invariant under adding local ancilla  in arbitrary state at Alice or Bob's site
\be
f(\rho_{AB}\ot\sigma_X)= f(\rho_{AB})
\ee
for $X=A',B'$. 
\item $f$ is affine on locally orthogonal states i.e. 
\label{affine}
\be
f(\sum_ip_i \rho_{AB}^i \ot |i\>\<i|)=\sum_ip_i
f( \rho_{AB}^i \ot |i\>_X\<i|)
\ee
for $X=A',B'$, where $|i\>$ are local, orthogonal flags.
\eee
\end{theorem}
{\bf Remark.} Since $f$ is assumed to be convex, in condition \ref{affine})
it is enough to check inequality in one direction.
One can formulate the conditions in a more elegant way as follows 
\begin{theorem}
\label{thm:2}
For a convex function $f$ does not increase under  LOCC if and only if 

\noindent{\rm [LUI]} $f$ satisfies  local unitary invariance (LUI)
\be
f(U_A\ot U_B \rho_{AB} U_A^\dagger\ot U_B^\dagger)= f(\rho_{AB})
\ee
\noindent {\rm [FLAGS]} $f$ satisfies 
\be
f(\sum_ip_i \rho_{AB}^i \ot |i\>_X\<i|)=\sum_ip_i
f( \rho_{AB}^i)
\label{eq:flags}
\ee
for $X=A',B'$ 
where $|i\>$ are local, orthogonal flags.
\end{theorem}
The condition FLAGS
is very intuitive: if we have a mixture of states with local flags,
then it is very reasonable to assume that the mixture has entanglement 
equal to average entanglement of the states. 

{\bf Remark.} In the proof we will not use the fact that the 
system is bipartite, so that the theorem holds also for multiparty systems.
It is also worth mentioning that the condition LUI
is usually  immediate to verify, so that for convex functions 
monotonicity is in a sense reduced just to the condition FLAGS.

{\bf Proof of equivalence.}
Let us argue that the theorems are equivalent. 
Since the condition LUI is a restatement of condition 1) of Theorem \ref{thm:1},
we need to show that FLAGS is equivalent to conditions 2) and 3).
First let us see that the condition FLAGS implies condition 2) 
of Theorem \ref{thm:1}. To this end, we consider spectral decomposition of 
the state $\sigma_X=\sum_k q_k |\phi_k\>\<\phi_k|$. 
Now in condition FLAGS, we take probabilities to be equal to $q_k$,
flags to be $\phi_k$ and the states $\rho_{AB}^i=\rho_{AB}$,
and get the condition 2). Now let us derive condition 3).
To get it from FLAGS we need the following equality 
\be
f(\rho_{AB}^i)=f(\rho_{AB}^i\ot 
|i\>_X\<i|)
\label{eq:fl}
\ee
This however is a consequence of condition 2), which as we have just shown 
follows from FLAGS.  

Now we need to prove that conditions 2) and 3) imply FLAGS. 
Obviously condition 2) implies (\ref{eq:fl}),
which together with 3) gives FLAGS. This ends the proof of equivalence 
of theorems. \blacksquare

{\bf Proof of Theorem \ref{thm:1}.} 
($"\Rightarrow" $)We will  first show that from conditions 1-3 we obtain monotonicity. 
Since our function is assumed to be convex, and by conditions 1) and 2) 
is nonincreasing under local 
unitaries and nonincreasing under adding local ancilla
(we also assume that it is {\it nondecreasing} under adding local ancilla)
it remains to show that it satisfies conditions b) and d) of Theorem \ref{thm:vidal}.
Let us first show that $f$ does not increase under local partial 
trace. For definiteness let the subsystem 
be at Bob's site. The initial state is $\rho_{ABB'}$,
the final state is $\rho_{AB}$.
Let us note that removing  subsystem can 
be performed in two stages: first one can apply suitable random unitaries,
to turn state into $\rho_{AB}\ot \tau_{B'}$ (where 
$\tau_{B'}$ is maximally mixed state on the subsystem $B'$) and then remove 
completely decoupled subsystem $B'$. In first stage $f$ will 
not increase, because of convexity and LUI. In second stage 
it will not increase because of condition 2). 

Now we will prove that $f$ does not increase under local measurement (e.g. on Bob's 
site). Consider a local measurement, that from $\rho_{AB}$ 
produces ensemble $p_i,\sigma_{AB}^i$. Note that it can be performed in 
the following way: 
\bei
\item Bob attaches local ancilla $B'$  in standard state $|0\>$
\item Bob applies unitary operation $U_{BB'}$
\item Bob measures ancilla by von Neumann measurement on system $B'$
\item Bob discards ancilla, and tells the outcome to Alice
\eei
Given the outcome $i$ of the measurement on $B'$, the state on $AB$ 
collapses to the state $\sigma_{AB}^i$. Now, we will replace the 
last stage with {\it dephasing} in basis
$|i\>_{B'}$ in which measurement was performed.
The dephasing can be performed by applying at random some 
unitary operations on system $B'$. The resulting 
total state will be then 
\be
\sum_ip_i \sigma_{AB}^i\ot |i\>_{B'}\<i|
\ee
Note that during operations leading to this state, $f$ didn't increase. 
Indeed, the operations are either mixing, or local unitaries, 
or adding ancilla. In first case function does not increase 
because of convexity, in second by condition 1) (LUI), in third one by condition 2).
Thus 
\be
f(\rho_{AB})\geq \sum_ip_if(\sigma_{AB}^i\ot |i\>_{B'}\<i|)
\ee
Now using 2), we get 
\be
f(\rho_{AB})\geq \sum_ip_if(\sigma_{AB}^i)
\ee
which is precisely  the condition d). 

($"\Leftarrow" $) Let us now prove the converse. Thus we assume that a 
function is convex and it is LOCC monotone. The condition LUI is 
is simply condition c), and the conditions  a) and b)  together 
imply condition 2), so that it is enough to show that 3) is implied. 
The inequality "$\leq$" in the condition follows from the fact that 
function is convex. Let us now prove inequality "$\geq$".
This follows immediately from condition d), if one takes  the (incomplete) measurement
to be measurement of the flags. This ends the proof. \blacksquare

We will illustrate our theorem by several examples.

{\it Example 1.} A well known measure of entanglement is 
{\it negativity} \cite{ZyczkowskiHSP-vol,Vidal-Werner} given by 
\be
E_N(\rho)=\|\rho^{T_A}\|=\|\rho^{T_B}\|
\ee
where $T_X$ is partial transpose performed on subsystem $X$ \cite{Peres96},
and $\|(\cdot) \|$ is trace norm. It was shown to be 
LOCC monotone in the sense of definition \ref{def:mon} in \cite{Vidal-Werner}.
A simple proof of monotonicity in the sense of  (\ref{eq:monweak})
was provided in \cite{irrev,irrev-errata}. 
Now, using our result, we are able to provide equally simple proof 
of stronger monotonicity of Def. \ref{def:mon}.
Of course the function $E_N$ is convex, because  partial  transpose 
is linear, and norm is convex. We have to prove the conditions LUI and 
FLAGS. The condition LUI follows from the fact that partial transpose 
in a sense commutes with local unitaries. Namely for unitaries 
$U_A$ and $W_B$ we have  

\be
U_A \ot W_B \rho_{AB}^{T_A}U_A^\dagger  \ot W_B^\dagger = 
(U_A \ot \tilde W_B \rho_{AB}U_A^\dagger  \ot \tilde W_B^\dagger )^{T_A}
\ee
where $\tilde W_B$ is again some unitary.
Let us now pass to condition FLAGS. 
First of all it is easy to see that for operators $A_i$ of disjoint 
supports we have 
\be
\|\sum_i A_i\|=\sum_i \|A_i\|
\ee
We will now take $A_i=p_i(\rho_{AB}^i)^{T_A} \ot |i\>B'\<i|$.
Because of orthogonal flags, the operators have disjoint supports,
so that we get 
\ben
&&\biggl\|\sum_i p_i(\rho_{AB}^i)^{T_A} \ot |i\>_B'\<i|\biggr\| = \\ \nonumber
&&=\sum_ip_i \bigl\|(\rho_{AB}^i)^{T_A} \ot |i\>_B'\<i|\bigr\|=\\ \nonumber
 && =\sum_ip_i \|(\rho_{AB}^i)^{T_A} \|
\een
the last inequality follows from the property of trace norm 
$\|A\ot B\|=\|A\| \ot \|B\|$.

{\it Example 2.} Consider relative entropy of entanglement 
\cite{PlenioVedral1998} given by 
\be
E_R(\rho)=\inf_{\sigma\in S} S(\rho|\sigma)
\ee
where $S$ is the set of separable states, and 
$S(\rho|\sigma)=\tr \rho\log \rho-\tr \rho\log \sigma$.
The proof of weaker monotonicity is immediate from 
definition of the measure \cite{PlenioVedral1998}. 
The proof of stronger monotonicity 
is somewhat more involved. However, due to double convexity 
of relative entropy, we know that relative entropy of entanglement 
is convex. Then we can apply our criteria. 
LUI follows immediately from invariance of set $S$ under local 
unitary operations. 
Let us prove that also FLAGS is satisfied. Again, we have to prove 
inequality "$\geq$". To see it, consider arbitrary  separable states $\sigma_{ABB'}$. 
Since relative entropy does not increase under dephasing, 
and set of separable states is closed under local dephasing, 
we can dephase the state on subsystem $B'$ in basis given by flags $|i\>$ 
and the obtained state can be only better candidate for infimum 
on the lhs of (\ref{eq:flags}). The new state is of the form 
$\sum_i p_i\sigma^i_{AB}\ot |i\>\<i|$ where $\sigma_{AB}^i$  
are again separable states. Because of orthogonality of flags we have 
\ben
&&S(\sum_ip_i\rho^i_{AB}\ot |i\>_{B'}\<i| \big| \sum_i p_i\sigma^i_{AB}\ot |i\>_{B'}\<i|)\\ \nonumber
&&=\sum_ip_iS(\rho^i_{AB}|\sigma^i_{AB}) 
\een
Thus for any candidate for infimum of left-hand-side,
we get a candidate for infimum of right-hand-side, 
which proves the inequality. 

To summarize, we have shown, that for a convex function, 
the LOCC monotonicity is equivalent to two 
simple conditions, local unitary invariance and 
the condition called FLAGS, which roughly speaking means, 
that the measure should not go down,
if we mix states that can be locally distinguished without disturbance. 
It is rather obvious that that the condition is necessary for 
monotonicity. However it might be surprising that 
for convex functions satisfying LUI it is also sufficient. It is also interesting, 
that the conditions are not inequalities, as might 
be expected from the nature of monotonicity. The only 
inequality is hidden in the condition of convexity. 

Since the condition FLAGS turned out to be so powerful 
for functions that are convex and satisfy local unitary invariance,
we think it can be also considered on its own, 
as an important property in the context of distant lab paradigm. 
However we should remember that in presence of activation effects 
\cite{activation,ShorST2001}
we would expect that it does not hold for distillable entanglement.
(The reason is the same for which the latter measure 
is expected to be nonconvex.)
Finally, we hope that the conditions we derived 
will simplify proofs of monotonicity for new entanglement 
measures, in particular, it may help to determine, 
if candidates for entanglement measures proposed in \cite{dual}
satisfy LOCC monotonicity (though we do not know 
whether the proposed functions are convex). 

We also hope, that the result may increase our understanding 
of what is actually hidden behind the postulate of nonincreasing under local 
operations and classical communication. 

The author would like to thank Ryszard Horodecki and Barbara Synak for discussion. 
This paper has been supported by the Polish Ministry of Scientific
Research and Information Technology under the (solicited) grant No.
PBZ-MIN-008/P03/2003 and by EC grants RESQ, Contract No. IST-2001-37559 
and  QUPRODIS, Contract No. IST-2001-38877.

\bibliography{refmich}
\bibliographystyle{apsrev}

\end{document}